# Magnetic properties and EPR spectrum of CsPr(MoO$_4$)$_2$

M. Kobets, K. Dergachev, E. Khatsko

E-mail: khatsko@ilt.kharkov.ua

*Institute for Low Temperature Physics National Academy of Science of Ukraine*

*47 Lenin Ave. 61103 Kharkov, Ukraine*

The temperature dependencies of magnetic susceptibility in temperature range 4.2-250 K along the principal magnetic axes as well as frequency-field dependence EPR spectrum at helium temperature in basic plane was studied. It is found that ground state of CsPr(MoO$_4$)$_2$, is the quasi-doublet. The experimental values of g-factors $g_c$=1.54±0.02; $g_a$<0.1; $g_b$<0.1 and energy gap $\Delta \sim 0{,}2$ cm$^{-1}$ was determined

Cesium-Praseodymium double molybdate CsPr(MoO$_4$)$_2$ belong to a class of double alkali-rare earth molybdates. These compounds are low symmetric magnetodielectrics with substantial magnetic moment in ground state. In such a crystal a low symmetry causes strong anisotropy of the all properties and its competition.

In particular, significant single ion anisotropy energy results in anisotropy of g-factor of the ground state as well as exited one, and anisotropy of spin-spin interaction. This result in to some distinctive peculiarities do not characteristic to the high symmetric crystals.

The foregoing determinates actuality and aim of investigation: study of the magnetic and resonant properties of CsPr(MoO$_4$)$_2$ compound.

The CsPr(MoO$_4$)$_2$ is characterized by rhombic symmetry space group $D_{2h}^3$ (P$_{ccm}$) [1]. The cell parameters are: $a$ = 9,56 Å; $b$ = 8,26 Å; $c$ = 5,14 Å, there are two formula units in the unit cell. The crystalline structure fragment of CsPr(MoO$_4$)$_2$ is represented on Fig.1. As can be seen, the structure is arranged in such a way that the layers of rare earth magnetic ions are separate by layers of alkali diamagnetic metal. This result in two-dimensional magnetic structure.

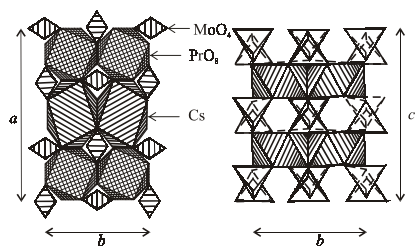

**Fig.1 The fragment of the CsPr(MoO4)2 crystalline structure**

The typical property of the rare earth compounds is a presence of four molecules in unit cell (two atoms along

*b*-axis and two in nonequivalent layers along *a*-axis). In this compound the rare earth layers are identical and situated against each other, so elementary cell of CsPr(MoO$_4$)$_2$ contain only two magnetic ions (Fig.2).

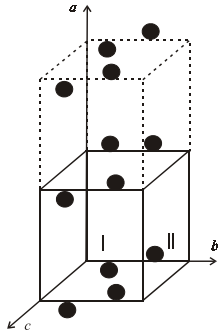

**Fig.2 The unit cell of CsPr(MoO4)2. By numbers the Pr$^{3+}$ ions in the same cell are marked.**

It is known that individual properties of magnetic ions are defined by crystal field potential and spin-orbit interaction. These properties play important role in forming of magnetic properties of magnetically concentrated crystals in paramagnetic region.

The temperature dependencies of magnetic susceptibility CsPr(MoO$_4$)$_2$ along three principal magnetic axes were investigated in [2]. All components of susceptibility are different and intersection of $\chi_x$(T) and $\chi_z$(T) curves was found. The nature of anisotropy of CsPr(MoO$_4$)$_2$ show that the symmetry of surrounding of Pr$^{3+}$ ion is not higher than rhombic.

To verify our sample we perform measurement of magnetic susceptibility in temperature range 2-250 K along the principal magnetic axes. These results are represented in Fig.3. The data corresponds to [2].

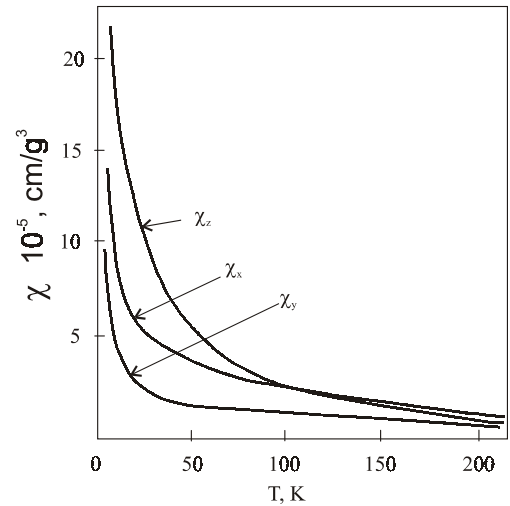

**Fig.3.The temperature dependence of the magnetic susceptidiliey of CsPr(MoO$_4$)$_2$ single crystal along the principal magnetic axes.**

The ground state of Pr$^{3+}$ ion is the term $^3H_4$, $4f^2$, S=1, L=5, J=4. This trivalent ion is non Kramers, as it have an even numbers of electrons. That means that at low temperature depending on the values and the ratio of crystal field constants the degeneration can be completely lifted and nonmagnetic ground state (singlet or quasi doublet) will be arise. This is the case of low symmetric components of rhombic crystalline field.

The magnetic properties of rare earth Pr$^{3+}$ ion in CsPr(MoO$_4$)$_2$ are defined by energy scheme of splitting of energy levels in crystalline field of matrices. The energy scheme sufficiently depends on a small crystalline field deviation. To determinate the energy scheme of ground term $^3H_4$ splitting of $Pr^{3+}$ ion in single crystal $CsPr(MoO_4)_2$ splitting the optical absorption spectra was studied in 1-2.5 mcm wave length diapason [2]. The local symmetry of $Pr^{3+}$ - $C_2$ is low enough, so multiplet of the $Pr^{3+}$ ion with even number of electron ($^3H_4$) must split by rhombic crystalline field to 9 energy levels. But in [2] the authors find only 8 optical transitions (14 cm$^{-1}$ to second exite state) they suppose, that ground state of $Pr^{3+}$ ion is quasi doublet, which they can not resolve. If it is so, in common diapason of magnetic fields and frequencies it must be observed a paramagnetic absorption (EPR).

Such an absorption was detected in $CsPr(MoO_4)_2$ at helium temperature and external magnetic field orientation along *c*-axis on frequency 78.1 GHz.

For detailed investigation of EPR spectrum in wide interval of frequencies 15–120 GHz the field-frequency dependencies was studied at helium temperatures.

The frequency dependences of EPR spectrum along *c*-axis was studied. In two another direction *a* and *b,* the g factors were found near zero ($g_a<0.1$; $g_b<0.1$), so this compound is good example of Ising magnet. Along *c*-axis the single absorption line is observed in all frequency diapason. This fact correspond the crystalline structure of $CsPr(MoO_4)_2$ when the unit cell contain two $Pr^{3+}$ ions connected one with other by inversion.

Resonant absorption in this compound can be observed only because of distortion of a crystalline field, which mixed two conjugate state. As result a small splitting $\Delta_0$ appears. The absorption line have asymmetric shape and its intensity is maximum when **H** || **h,** and $g_\perp=0$ as is observed in experiment.

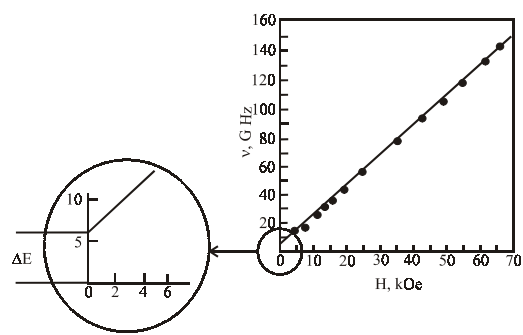

**Fig.4. Frequency-field dependences of EPR spectrum of $CsPr(MoO_4)_2$ compound H||h||*c*.** T=4.2 K

Fig.4 represent the frequency-field dependences of EPT spectrum along *c*-axis at **H**||**h** polarization. It is seen that ground state is spited and the value of gap is $\Delta \sim 0{,}2$ cm$^{-1}$

The resonant frequency $\omega/\gamma$ dependence versus value and direction of external magnetic field **H** is defined by expression:

$\omega/\gamma = (\Delta^2 + \mu_B^2 g^2 H^2 \cos^2\varphi)^{1/2}$. So, frequency-field investigation of EPR spectrum of $Pr^{3+}$ ion in $CsPr(MoO_4)_2$ allow to supplement and restore the experimental scheme of multiplet ($^3H_4$) of $Pr^{3+}$ ion with even number of electrons; estimate the energy gap in quasi doublet $\Delta \sim 0{,}2$ cm$^{-1}$ and determine the g-factors values along crystallographic axes $g_c = 1{,}54 \pm 0{,}02$; $g_a < 0.1$; $g_b < 0.1$.